\documentclass[11pt]{article}%
\usepackage{amsmath}
\usepackage{graphicx}%
\usepackage{amsfonts}%
\usepackage{amssymb}

\newcommand{\eqnb}{\begin{equation}}
\newcommand{\eqne}{\end{equation}}

\newtheorem{The}{Theorem}

\newtheorem{Cor}[The]{Corollary}

\newtheorem{Rem}{Remark}

\begin{document}

\title{Markov Processes in Blockchain Systems}
\author{Quan-Lin Li$^{a}$, Jing-Yu Ma$^{b}$, Yan-Xia Chang$^{c}$, Fan-Qi Ma$^{b}$,
Hai-Bo Yu$^{a}$\\$^{a}$School of Economics and Management,\\Beijing University of Technology, Beijing 100124, China\\$^{b}$School of Economics and Management,\\Yanshan University, Qinhuangdao 066004, China\\$^{c}$School of Science, Yanshan University, Qinhuangdao 066004, China}
\maketitle

\begin{abstract}
In this paper, we develop a more general framework of block-structured Markov
processes in the queueing study of blockchain systems, which can provide
analysis both for the stationary performance measures and for the sojourn
times of any transaction or block. Note that an original aim of this paper is
to generalize the two-stage batch-service queueing model studied in Li et al.
\cite{Li:2018} both ``from exponential to phase-type" service times and ``from
Poisson to MAP" transaction arrivals. In general, the MAP transaction arrivals
and the two stages of PH service times make our blockchain queue more suitable
to various practical conditions of blockchain systems with crucial factors,
for example, the mining processes, the block-generations, the
blockchain-building and so forth. For such a more general blockchain queueing
model, we focus on two basic research aspects: (1) By using the
matrix-geometric solution, we first obtain a sufficient stable condition of
the blockchain system. Then we provide simple expressions for the average
stationary number of transactions in the queueing waiting room, and the
average stationary number of transactions in the block. (2) However, comparing
with Li et al. \cite{Li:2018}, analysis of the transaction-confirmation time
becomes very difficult and challenging due to the complicated blockchain
structure. To overcome the difficulties, we develop a computational technique
of the first passage times by means of both the PH distributions of infinite
sizes and the $RG$-factorizations. Finally, we hope that the methodology and
results given in this paper will open a new avenue to queueing analysis of
more general blockchain systems in practice, and can motivate a series of
promising future research on development of blockchain technologies.

\vskip               0.5cm


\end{abstract}

\section{Introduction}

\textbf{Background and Motivation. }Blockchain is one of the most popular
issues discussed extensively in recent years, and it has already changed
people's lifestyle in some real areas due to its great impact on finance,
business, industry, transportation, heathcare and so forth. Since the
introduction of Bitcoin by Nakamoto \cite{Nak:2008}, blockchain technologies
has obtained many important advances in basic theory and real applications up
to now. Readers may refer to, for example, excellent books by Wattenhofer
\cite{Wat:2016}, Prusty \cite{Pru:2017}, Drescher \cite{Dre:2017}, Bashir
\cite{Bas:2018} and Parker \cite{Par:2018}; and survey papers by Zheng et al.
\cite{Zhe:2017}, Constantinides et al. \cite{Con£º2018}, Yli-Huumo et al.
\cite{Yli:2016}, Lindman et al. \cite{Lind:2017} and Risius and Spohrer
\cite{Ris:2017}.

It may be necessary and useful to further remark several important directions
and key research as follows: \textbf{(1)} Smart contracts by Reed
\cite{Ree:2016}, Bartoletti and Pompianu \cite{Bar:2017}, Alharby and van
Moorsel \cite{Alh:2017} and Magazzeni et al. \cite{Mag:2017}. \textbf{(2)}
Ethereum by Diedrich \cite{Die:2016}, Dannen \cite{Dan:2017}, Atzei et al.
\cite{Atz:2017} and Antonopoulos and Wood \cite{Ant:2018}. \textbf{(3)}
Consensus mechanisms by Wang et al. \cite{Wan:2018}, Debus \cite{Deb:2017},
Pass et al. \cite{Pas:2017}, Pass and Shi \cite{PasS:2017} and Cachin and
Vukoli\'{c} \cite{Cac:2017}. \textbf{(4)} Blockchain security by Karame and
Androulaki \cite{Kar:2016}, Lin and Liao \cite{Lin:2017} and Joshi et al.
\cite{Jos:2018}. \textbf{(5)} Blockchain economics by Swan \cite{Swa:2015},
Catalini and Gans \cite{Cat£º2016}, Davidson et al. \cite{Dav:2016}, Bheemaiah
\cite{Bhe:2017}, Becket al. \cite{Bec:2018} and Abadi and Brunnermeier
\cite{Aba:2018}. In addition, there are still some important topics including
the mining management, the double spending, PoW, PoS, PBFT, withholding
attacks, pegged sidechains and so on, and also their investigations may be
well understood from the references listed above.

Recently, blockchain has become widely adopted in many real applications.
Readers may refer to, for example, Foroglou and Tsilidou \cite{For:2015},
Bahga and Madisetti \cite{Bahg:2017} and Xu et al. \cite{Xu:2019}. At the same
time, we also provide a detailed observation on some specific perspectives,
for instance, \textbf{(1) }blockchain finance by Tsai et al. \cite{Tsa:2016},
Nguyen \cite{Ngu:2016}, Tapscott and Tapscott \cite{Tap:2017}, Treleaven et
al. \cite{Tre:2017} and Casey et al. \cite{Cas:2018}; \textbf{(2) }blockchain
business by Mougayar \cite{Mou:2016}, Morabito \cite{Mor:2017}, Fleming
\cite{Fle:2017}, Beck et al. \cite{Bec:2017}, Nowi\'{n}ski and Kozma
\cite{Now:2017} and Mendling et al. \cite{Men:2018}; \textbf{(3)} supply
chains under blockchain by Hofmann et al. \cite{Hof:2017}, Korpela et al.
\cite{Kor:2017}, Kim and Laskowski \cite{Kim:2018}, Saberi et al.
\cite{Sab:2018}, Petersen et al. \cite{Pet:2018}, Sternberg and Baruffaldi
\cite{Ste:2018} and Dujak and Sajter \cite{Duj:2019}; \textbf{(4)} internet of
things under blockchain by Conoscenti et al. \cite{Con:2016}, Bahga and
Madisetti \cite{Bah:2016}, Dorri et al. \cite{Dor:2016}, Christidis and
Devetsikiotis \cite{Chr:2016} and Zhang and Wen \cite{Zhan:2017}; \textbf{(5)}
sharing economy under blockchain by Huckle et al. \cite{Huc:2016}, Hawlitschek
et al. \cite{Haw:2018}, De Filippi \cite{De:2017} and Pazaitis et al.
\cite{Paz:2017}; \textbf{(6)} healthcare under blockchain by Mettler
\cite{Met:2016}, Rabah \cite{Rab:2017}, Griggs et al. \cite{Gri:2018} and Wang
et al. \cite{Wang:2018}; \textbf{(7)} energy under blockchain by Oh et al.
\cite{Oh:2017}, Aitzhan and Svetinovic \cite{Aiz:2018}, Noor et al.
\cite{Noo:2018} and Wu and Tran \cite{Wu:2018}.

Based on the above discussions, whether it is theoretical research or real
applications, we always hope to know how performance of the blockchain system
are obtained, and wheth there is still some room to be able to further improve
performance of the blockchain system. Based on this, it is a key to find
solution of such a performance issue in the study of blockchain systems. Thus
we need to provide mathemtical modeling and analysis for blockchain
performance evaluation by means of, for example, Markov processes, Markov
decision processes, queueing networks, Petri networks, game models and so on.
Unfortunately, so far only a little work has been on performance modeling of
blockchain systems. Therefore, this motivates us in this paper to develop
Markov processes and queueing models for a more general blockchain system. We
hope that the methodology and results given in this paper will open a new
avenue to Markov processes of blockchain systems, and can motivate a series of
promising future research on development of blockchain technologies.

\textbf{Related Work. }Now, we provide several different classes of related
work for Markov processes in blockchain systems, for example, queueing models,
Markov processes, Markov decision processes, random walks, fluid limit and so on.

\textit{Queueing models}: To use queueing theory to model a blockchain system,
we need to observer some key factors, for example, transaction arrivals,
block-generation, block size, transaction fee, mining pools, mining reward,
solving difficulty of crypto mathematical puzzle, blockchain-building,
throughput and so forth. As sketched in Figure 1, we design a two stage,
Service-In-Random-Order and batch service queueing system by means of two
stages of asynchronous processes: Block-generation and blockchain-building. Li
et al. \cite{Li:2018} is the first one to provide a detailed analysis for such
a blockchain queue by means of the matrix-geometric solution. Kasahara and
Kawahara \cite{Kas:2016} and Kawase and Kasahara \cite{Kaw:2017} discussed the
blockchain queue with general service times through an incompletely solving
idea for dealing with an interesting open problem. In addition, they also gave
some useful numerical experiments for performance observation. Ricci et al.
\cite{Ric:2019} proposed a framework encompassing machine learning and a
queueing model, which is used to identify which transactions will be
confirmed, and to characterize the confirmation time of confirmed
transactions. Memon et al. \cite{Mem:2019} proposed a simulation model for the
blockchain systems by means of queuing theory.

Bowden et al. \cite{Bow:2018} discussed time-inhomogeneous behavior of the
block arrivals in the bitcoin blockchain because the block-generation process
is influenced by multiple key factors such as the solving difficulty level of
crypto mathematical puzzle, transaction fee, mining reward, mining pools and
so on. Papadis et al. \cite{Pap:2018} applied the time-inhomogeneous block
arrivals to set up some Markov processes to study evolution and dynamics of
blockchain networks, and discussed key blockchain characteristics such as the
number of miners, the hashing power (block completion rates), block
dissemination delays, and block confirmation rules. Further, Jourdan et al.
\cite{Jou:2018} proposed a probabilistic model of the bitcoin blockchain by
means of a transaction and block graph, formulated some conditional
dependencies induced by the bitcoin protocol at the block level. Based on
this, it is clear that when the block-generation arrivals are a
time-inhomogeneous Poisson process, we believe that the blockchain queue
analyzed in this paper will become very difficult and challenging, thus it
will be an interesting topic in our future study.

\textit{Markov processes:} To evaluate performance of a blockchain system,
Markov processes are a basic mathematical tool, e.g., see Bolch et al.
\cite{Bol:2006} for more details. As an early key work to apply Markov
processes to blockchain performance issues, Eyal and Sirer \cite{Eya:2018}
established a simple Markov process to analyze the vulnerability of Nakamoto
Protocols through studying the block-forking behavior of blockchain. Note that
some selfish miners may get higher payoffs by violating the information
propagation protocols and postponing their mined blocks so that such selfish
miners exploits the inherent block forking phenomenon of Nakamoto protocols.
Nayak et al. \cite{Nya:2016} extended the work by Eyal and Sirer
\cite{Eya:2018} through introducing a new mining strategy: Stubborn mining
strategy. They used three improved Markov processes to further study the
stubborn mining strategy and two extensions: the Equal-Fork Stubborn (EFS) and
the Trail Stubborn (TS) mining strategies. Carlsten \cite{Car:2016} used the
Markov process to study the impact of transaction fees on the selfish mining
strategies in the bitcoin network. G\"{o}bel et al. \cite{Gob:2016} further
considered the mining competition between a selfish mining pool and the honest
community by means of a two-dimensional Markov process, in which they extended
the Markov model of selfish mining by considering the propagation delay
between the selfish mining pool and the honest community.

Kiffer and Rajaraman \cite{Kif:2018} provided a simple framework of Markov
processes for analyzing consistency properties of the blockchain protocols,
and used some numerical experiments to check the consensus bounds for network
delay parameters and adversarial computing percentages. Huang et al.
\cite{Hua:2019} set up a Markov process with an absorbing state to analyze
performance measures of the Raft consensus algorithm for a private blockchain.

\textit{Markov decision processes:} Note that the selfish miner may adopt
different mining policies to release some blocks under the longest-chain rule
for controling the block-forking structure, thus it is interesting to find an
optimal mining policy in the blockchain system. To do this, Sapirshtein et al.
\cite{Sap:2016}, Sompolinsky and Zohar \cite{Som:2016} and Gervais et al.
\cite{Ger:2016} applied the Markov decision processes to find the optimal
selfish-mining strategy, in which four actions: adopt, override, match and
wait, are introduced in order to control the state transitions of the Markov
decision process.

\textit{Random walks:} Goffard \cite{Gof:2019} proposed a random walk method
to study the double-spending attack problems in the blockchin system, and
focused on how to evaluate the probability of the double-spending attack ever
being successful. Jang and Lee \cite{Jan:2019} discussed profitability of the
double-spending attacks in the blockchain systems through using the random
walk of two independent Poisson counting processes.

\textit{Fluid limit: }Frolkova and Mandjes \cite{Fro:2019} considered a
bitcoin-inspired infinite-server model with a random fluid limit. King
\cite{Kin:2019} developed the fluid limit of a random graph model to discuss
the shared ledger and the distributed ledger technologies in the blockchain systems.

\textbf{Contributions. }The main contributions of this paper are twofold. The
first contribution is to develop a more general framework of block-structured
Markov processes in the study of blockchain systems. We design a two stage,
Service-In-Random-Order and batch service queueing system, whose original aim
is to generalize the blockchain queue studied in Li et al. \cite{Li:2018} both
``from exponential to phase-type'' service times and ``from Poisson to MAP''
transaction arrivals. Note that the transaction MAP arrivals and two stages of
PH service times make our new blockchain queueing model more suitable to
various practical conditions of blockchain systems. By using the
matrix-geometric solution, we obtain a sufficient stable condition of the more
general blockchain system, and provide simple expressions for two key
performance measures: The average stationary number of transactions in the
queueing waiting room, the average stationary number of transactions in the block.

The second contribution of this paper is to provide an effective method for
computing the average transaction-confirmation time of any transaction in a
more general blockchain system. In general, it is always very difficult and
challenging to analyze the transaction-confirmation time in the blockchain
system with MAP inputs and PH service times, because the service discipline of
the blockchain system is new from two key points: (1) The ``block service'' is
a class of batch service; and (2) some transactions are chosen into a block
through Service-In-Random-Order. In addition, the MAP inputs and PH service
times also make analysis of the blockchain queue more complicated. To study
the transaction-confirmation time, we set up a Markov process with an
absorbing state (see Figure 4) according to the blockchain system (see Figures
1 and 2). Based on this, we show that the transaction-confirmation time of any
transaction is the first passage time of the Markov process with an absorbing
state, hence we can discuss the transaction-confirmation time (or the first
passage time) by means of both the PH distributions of infinite sizes and the
$RG$-factorizations. Further, we propose an effective algorithm for computing
the average transaction-confirmation time of any transaction. We hope that our
approach given in this paper can be applicable to deal with the
transaction-confirmation times in more general blockchain systems.

The structure of this paper is organized as follows. Section 2 describes a two
stage, Service-In-Random-Order and batch service queueing system, where the
transactions arrive at the blockchain system according to a Markovian arrival
process (MAP), the block-generation and blockchain-building times are all of
phase type (PH). Section 3 establishes a continuous-time Markov process of
GI/M/1 type, derives a sufficient stable condition of the blockchain system,
and expresses the stationary probability vector of the blockchain system by
means of the matrix-geometric solution. Section 4 provides simple expressions
for the average stationary number of transactions in the queueing waiting
room, the average stationary number of transactions in the block; and uses
some numerical examples to verify computability of our theoretical results. To
compute the average transaction-confirmation time of any transaction, Section
5 develops a computational technique of the first passage times by means of
both the PH distributions of infinite sizes and the $RG$-factorizations.
Finally, some concluding remarks are given in Section 6.

\section{Model Description}

In this section, from a more general point of view of blockchain, we design an
interesting and practical blockchain queueing system, where the transactions
arrive at the blockchain system according to a Markovian arrival process
(MAP), while the block-generation and blockchain-building times are all of
phase type (PH).

From a more practical background of blockchain, it is necessary to extend and
generalize the blockchain queueing model, given in Li et al. \cite{Li:2018},
to a more general case not only with non-Poisson transaction inputs but also
with non-exponential block-generation and blockchain-building times. Based on
this, we further abstract the block-generation and blockchain-building
processes as a queue of batch service, Service-In-Random-Order input and two
different service stages by means of the MAP and the PH distribution. Such a
blockchain queueing system is depicted in Figure 1.

\begin{figure}[th]
\centering            \includegraphics[width=12cm]{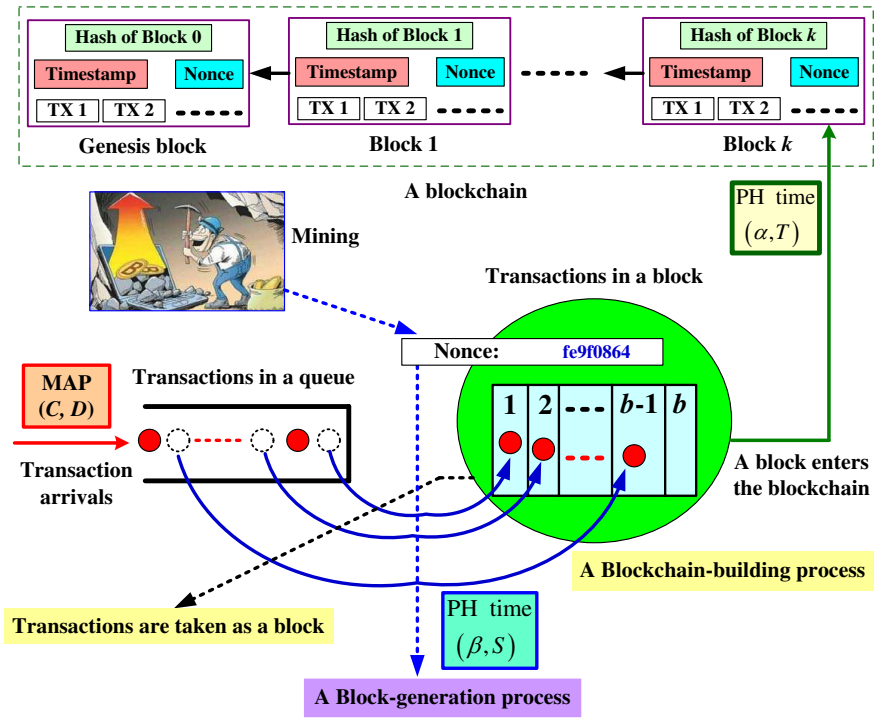}  \caption{A
blockchain queueing system under Markovian environment}%
\label{figure:Fig-1}%
\end{figure}

From Figure 1, now we provide some model descriptions as follows:

\textbf{Arrival process:} Transactions arrive at the blockchain system
according to a Markovian arrival process (MAP) with representation $\left(
C,D\right)  $ of order $m_{0}$, where the matrix $C+D$ is the infinitesimal
generator of an irreducible Markov process; $C$ indicates the state transition
rates that only the random environment changes without any transaction
arrival, $D$ denotes the arrival rates of transactions under the random
environment $C$; $\left(  C+D\right)  e=0$, and $e$ is a column vector of
suitable size in which each element is one. Obviously, the Markov process
$C+D$ with finite states is irreducible and positive recurrent. Let
$\mathbf{\omega}$ be the stationary probability vector of the Markov process
$C+D$, it is clear that $\omega\left(  C+D\right)  =0$ and $\omega e=1$. Also,
the stationary arrival rate of the MAP is given by $\lambda=\omega De$.

In addition, we assume that each arriving transaction must first enter a
queueing waiting room of infinite size. See the lower left part corner of
Figure 1.

\textbf{A block-generation process:} Each arriving transaction first needs to
queue in a waiting room. Then it is possibly chosen into a block of the
maximal size $b$. This is regarded as the first stage of service, called
\textit{block-generation} process. Note that the arriving transactions will be
continually chosen into the block until the block-generation process is over
under which a nonce is appended to the block by a mining winner. See the lower
middle part of Figure 1 for more details.

The block-generation time begins the initial epoch of a mining process until a
nonce of the block is found (i.e., the cryptographic mathematical puzzle is
solved for sending a nonce to the block), then the mining process is
terminated immediately. We assume that all the block-generation times are
i.i.d., and are of phase type with an irreducible representation $\left(
\beta,S\right)  $ of order $m_{2}$, where $\beta e=1$, the expected
blockchain-building time is given by $1/\mu_{2}=-\beta S^{-1}e$.

\textbf{The block-generation discipline:} A block can consist of some
transactions but at most $b$ transactions. Once the mining process begins, the
transactions are chosen into a block, in which they are not completely based
on the First Come First Service (FCFS) from the order of transaction arrivals.
In this case, several transactions in the back of this queue may also be
preferentially chosen into the block. When the block is formed, it will not
receive any new arriving transaction. See the lower middle part of Figure 1.

\textbf{A blockchain-building process:} Once the mining process is over, the
block with a group of transactions will be pegged to a blockchain. This is
regarded as the second stage of service due to the network latency, called
\textit{blockchain-building} process, see the lower right corner of Figure 1.
In addition, the upper part of Figure 1 also outlines the blockchain and the
internal structure of every block.

In the blockchain system, we assume that the blockchain-building times are
i.i.d, and have a common PH-distribution with an irreducible representation
$\left(  \alpha,T\right)  $ of order $m_{1}$, where $\alpha e=1$, and the
expected block-generation time is given by $1/\mu_{1}=-\alpha T^{-1}e$.

\textbf{The maximum block size:} To avoid the spam attacks, the maximum size
of each block is limited. We assume that there are at most $b$ transactions in
each block. If there are more than $b$ transactions in the queueing waiting
room, then the $b$ transactions are chosen into a full block so that those
redundant transactions are left in the queueing waiting room in order to set
up another possible block. In addition, the block size $b$ maximizes the batch
service ability in the blockchain system.

\textbf{Independence:} We assume that all the random variables defined above
are independent of each other.

\begin{Rem}
This paper is the first one to consider a blockchain system with non-Poisson
transaction arrivals (MAPs) and with non-exponential block-generation and
blockchain-building times (PH distributions), and it also provides a detailed
analysis for the blockchain queueing model by means of the block-structured
Markov processes and the $RG$-factorizations. However, so far analysis of the
blockchain queues with renewal arrival process or with general service time
distributions has still been an interesting open problem in queueing research
of blockchain systems.
\end{Rem}

\begin{Rem}
In the blockchain system, there are some key factors including the maximum
block size, mining reward, transaction fee, mining strategy, security of
blockchain and so on. Based on this, we may develop reward queueing models,
decision queueing models, and game queueing models in the study of blockchain
systems. Therefore, analysis for the key factors will be not only
theoretically necessary but also practically important in development of
blockchain technologies.
\end{Rem}

\section{A Markov Process of GI/M/1 Type}

In this section, to analyze the blockchain queueing system, we first establish
a continuous-time Markov process of GI/M/1 type. Then we derive a system
stable condition and express the stationary probability vector of the
blockchain queueing system by means of the matrix-geometric solution.

Let $N_{1}\left(  t\right)  ,N_{2}\left(  t\right)  ,I\left(  t\right)
,J_{1}\left(  t\right)  $ and$\ J_{2}\left(  t\right)  $ be the number of
transactions in the queueing waiting room, the number of transactions in the
block, the phase of the MAP, the phase of a blockchain-building PH time, and
the phase of a block-generation PH time at time $t$, respectively. We write
$\mathbf{X}=\left\{  \left(  N_{1}\left(  t\right)  ,N_{2}\left(  t\right)
,I\left(  t\right)  ,J_{1}\left(  t\right)  ,J_{2}\left(  t\right)  \right)
,t\geq0\right\}  $. Then it is easy to see that $\mathbf{X}$ is a
continuous-time Markov process with block structure whose state space is given
by%
\begin{align*}
\mathbf{\Omega}=  &  \left\{  \left(  0,0;i\right)  ,1\leq i\leq m_{0}\right\}
\\
&  \cup\left\{  \left(  0,l;i,j\right)  ,1\leq l\leq b,1\leq i\leq m_{0},1\leq
j\leq m_{1}\right\} \\
&  \cup\left\{  \left(  k,0;i,r\right)  ,k\geq1,1\leq i\leq m_{0},1\leq r\leq
m_{2}\right\} \\
&  \cup\left\{  \left(  k,l;i,j\right)  ,k\geq1,1\leq l\leq b,1\leq i\leq
m_{0},1\leq j\leq m_{1}\right\}  .
\end{align*}
From Figure 1, it is easy to set up the state transition relation of the
Markov process $\mathbf{X}$, see Figure 2 for more details. It is a key in
understanding of Figure 2 that there is a different transition between State
$\left(  k,0\right)  $ for the block-generation and State $\left(  k,l\right)
$ for the blockchain-building with $1\leq l\leq b$ because the
block-generation and blockchain-building processes can not simultaneously
exist at a time, and specifically, a block must first be generated, then it
can enter the blockchain-building process.

\begin{figure}[th]
\centering              \includegraphics[width=\textwidth]{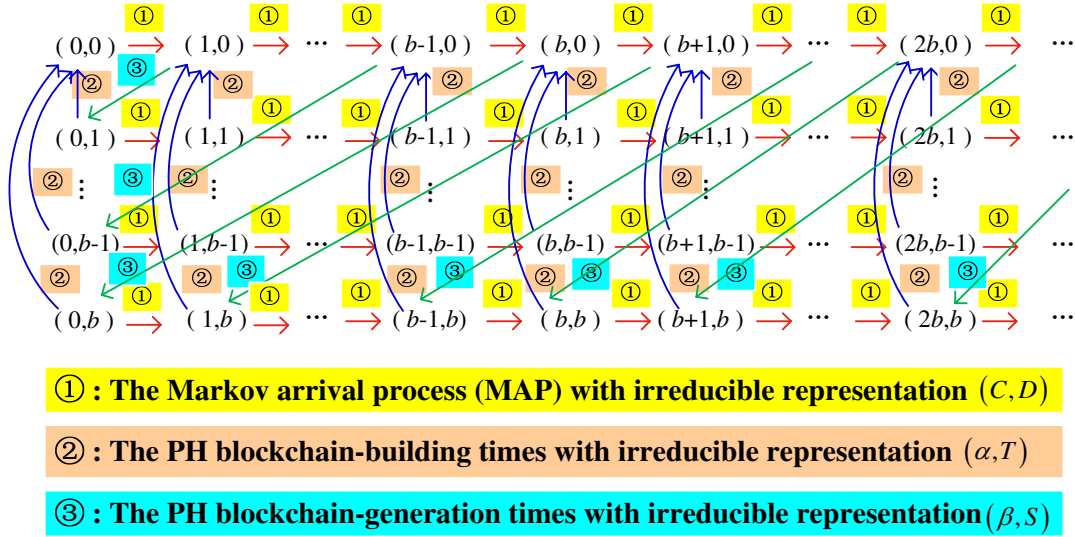}
\newline \caption{State transition relation of the Markov process}%
\label{figure:Fig-2}%
\end{figure}

By using Figure 2, the infinitesimal generator of the Markov process
$\mathbf{X}$ is given by
\begin{equation}
\mathbf{Q}=\left(
\begin{array}
[c]{ccccccccc}%
B_{1} & B_{0} &  &  &  &  &  &  & \\
B_{2} & A_{1} & A_{0} &  &  &  &  &  & \\
B_{3} &  & A_{1} & A_{0} &  &  &  &  & \\
\vdots &  &  & \ddots & \ddots &  &  &  & \\
B_{b+1} &  &  &  & A_{1} & A_{0} &  &  & \\
& A_{b+1} &  &  &  & A_{1} & A_{0} &  & \\
&  & A_{b+1} &  &  &  & A_{1} & A_{0} & \\
&  &  & \ddots &  &  &  & \ddots & \ddots
\end{array}
\right)  , \label{eq-2}%
\end{equation}
where $\otimes$ and $\oplus$ are the Kronecker product and the Kronecker sum
of two matrices, respectively,
\[
A_{0}=\left(
\begin{array}
[c]{cccc}%
D\otimes I &  &  & \\
& D\otimes I &  & \\
&  & \ddots & \\
&  &  & D\otimes I
\end{array}
\right)  ,\text{ }A_{1}=\left(
\begin{array}
[c]{cccc}%
C\oplus S &  &  & \\
I\otimes\left(  T^{0}\beta\right)  & C\oplus T &  & \\
\vdots &  & \ddots & \\
I\otimes\left(  T^{0}\beta\right)  &  &  & C\oplus T
\end{array}
\right)  ,
\]%
\[
A_{b+1}=\left(
\begin{array}
[c]{cccc}%
0 & \cdots & 0 & I\otimes\left(  S^{0}\alpha\right) \\
&  &  & \\
&  &  & \\
&  &  &
\end{array}
\right)  ,
\]
and%
\[
B_{0}=\left(
\begin{array}
[c]{cccc}%
D\otimes\beta &  &  & \\
& D\otimes I &  & \\
&  & \ddots & \\
&  &  & D\otimes I
\end{array}
\right)  ,\text{ }B_{1}=\left(
\begin{array}
[c]{cccc}%
C &  &  & \\
I\otimes T^{0} & C\oplus T &  & \\
\vdots &  & \ddots & \\
I\otimes T^{0} &  &  & C\oplus T
\end{array}
\right)  ,
\]%
\[
B_{2}=\left(
\begin{array}
[c]{ccccc}%
0 & I\otimes\left(  S^{0}\alpha\right)  & 0 & \cdots & 0\\
&  &  &  & \\
&  &  &  & \\
&  &  &  &
\end{array}
\right)  ,\text{ \ }\ldots,\text{ \ }B_{b+1}=\left(
\begin{array}
[c]{cccc}%
0 & \cdots & 0 & I\otimes\left(  S^{0}\alpha\right) \\
&  &  & \\
&  &  & \\
&  &  &
\end{array}
\right)  .
\]
Clearly, the continuous-time Markov process $\mathbf{X}$ is of GI/M/1 type.

Now, we use the mean drift method to discuss the system stable condition of
the continuous-time Markov process $\mathbf{X}$ of GI/M/1 type. Note that the
mean drift method for checking system stability is given a detailed
introduction in Chapter 3 of Li \cite{Li:2010}.

From Chapter 1 of Neuts \cite{Neu:1981} or Chapter 3 of Li \cite{Li:2010}, for
the Markov process of GI/M/1 type, we write%
\begin{align*}
\mathbf{A}  &  =A_{0}+A_{1}+A_{b+1}\\
&  =\left(
\begin{array}
[c]{ccccc}%
D\otimes I+C\oplus S &  &  &  & I\otimes\left(  S^{0}\alpha\right) \\
I\otimes\left(  T^{0}\beta\right)  & D\otimes I+C\oplus T &  &  & \\
\vdots &  & \ddots &  & \\
I\otimes\left(  T^{0}\beta\right)  &  &  & D\otimes I+C\oplus T & \\
I\otimes\left(  T^{0}\beta\right)  &  &  &  & D\otimes I+C\oplus T
\end{array}
\right)  .
\end{align*}
Clearly, the matrix $\mathbf{A}$ is the infinitesimal generator of an
irreducible, aperiodic and positive recurrent Markov process with two levels
(i.e., Levels $0$ and $b$), together with $b-1$ instantaneous levels (i.e.,
Levels $1,2,...,b-1$) which will vanish as the time $t$ goes to infinity. On
the other hand, such a special Markov process $\mathbf{A}$ will not influence
applications of the matrix-geometric solution because it is only related to
the mean drift method for establishing system stable conditions.

The following theorem discusses the invariant measure $\theta$ of the Markov
process $\mathbf{A}$, that is, the vector $\theta$ satisfies the system of
linear equations $\theta\mathbf{A}=0$ and $\theta e=1$.

\begin{The}
There exists the unique invariant measure $\theta=\left(  \theta_{0}%
,0,\ldots,0,\theta_{b}\right)  $ of the Markov process $\mathbf{A}$, where
$\left(  \theta_{0},\theta_{b}\right)  $ is the stationary probability vector
of the irreducible positive-recurrent Markov process whose infinitesimal
generator%
\[
\Re=\left(
\begin{array}
[c]{cc}%
D\otimes I+C\oplus S & I\otimes\left(  S^{0}\alpha\right) \\
I\otimes\left(  T^{0}\beta\right)  & D\otimes I+C\oplus T
\end{array}
\right)  .
\]
\end{The}

\textbf{Proof: }It follows from $\theta\mathbf{A}=0$ that
\begin{equation}
\theta_{1}\left(  D\otimes I+C\oplus S\right)  +\sum_{k=1}^{b-1}\theta
_{k}\left[  I\otimes\left(  T^{0}\beta\right)  \right]  +\theta_{b}\left[
I\otimes\left(  T^{0}\beta\right)  \right]  =0\mathbf{,} \label{Equa-1}%
\end{equation}%
\begin{equation}
\theta_{k}\left[  D\otimes I+C\oplus T\right]  =0,\text{ \ }1\leq k\leq b-1,
\label{Equa-2}%
\end{equation}%
\begin{equation}
\theta_{1}\left[  I\otimes\left(  S^{0}\alpha\right)  \right]  +\theta
_{b}\left(  D\otimes I+C\oplus T\right)  =0. \label{Equa-3}%
\end{equation}
For Equation (\ref{Equa-2}), note that%
\begin{align*}
D\otimes I+C\oplus T  &  =D\otimes I+C\otimes I+I\otimes T\\
&  =\left(  C+D\right)  \otimes I+I\otimes T\\
&  =\left(  C+D\right)  \oplus T,
\end{align*}
where $C+D$ is the infinitesimal generator of an irreducible and a
positive-recurrent Markov process, thus its eigenvalue of the maximal real
part is zero so that all the other eigenvalues have a negative real part;
while $T$, coming from the PH distribution with irreducible representation
$\left(  \alpha,T\right)  $, is invertible with the real part of each
eigenvalue be negative due to the fact that $Te\lvertneqq0$, and the matrix
$T$ has the properties that all diagonal elements are negative, and all
off-diagonal elements are nonnegative. Note that each eigenvalue of the matrix
$\left(  C+D\right)  \oplus T$ are the sum of an eigenvalue of the matrix
$C+D$ and an eigenvalue of the matrix $T$,\ thus each eigenvalue of the matrix
$\left(  C+D\right)  \oplus T$ has a negative real part (i.e., it is
non-zero). This shows that the matrix $\left(  C+D\right)  \oplus T$ is
invertible by means of det$\left(  \left(  C+D\right)  \oplus T\right)  \neq
0$, which is the product of all the eigenvalues of $\left(  C+D\right)  \oplus
T$. Hence, from Equation $\theta_{k}\left[  D\otimes I+C\oplus T\right]  =0$
for $1\leq k\leq b-1$, we obtain%
\[
\theta_{1}=\theta_{2}=\cdots=\theta_{b-1}=0.
\]
This gives
\[
\theta=\left(  \theta_{0},0,\ldots,0,\theta_{b}\right)  .
\]
It follows from (\ref{Equa-1}) and (\ref{Equa-3}) that%
\[
\left\{
\begin{array}
[c]{c}%
\theta_{0}\left(  D\otimes I+C\oplus S\right)  +\theta_{b}\left[
I\otimes\left(  T^{0}\beta\right)  \right]  =0\mathbf{,}\\
\theta_{0}\left[  I\otimes\left(  S^{0}\alpha\right)  \right]  +\theta
_{b}\left(  D\otimes I+C\oplus T\right)  =0\mathbf{.}%
\end{array}
\right.
\]
Thus we have%
\[
\left(  \theta_{0},\theta_{b}\right)  \left(
\begin{array}
[c]{cc}%
D\otimes I+C\oplus S & I\otimes\left(  S^{0}\alpha\right) \\
I\otimes\left(  T^{0}\beta\right)  & D\otimes I+C\oplus T
\end{array}
\right)  =\left(  0,0\right)  .
\]
Let%
\[
\Re=\left(
\begin{array}
[c]{cc}%
D\otimes I+C\oplus S & I\otimes\left(  S^{0}\alpha\right) \\
I\otimes\left(  T^{0}\beta\right)  & D\otimes I+C\oplus T
\end{array}
\right)  .
\]
Then the matrix $\Re$ is the infinitesimal generator of an irreducible
positive-recurrent Markov process. Thus the Markov process $\Re$ exists the
stationary probability vector $\left(  \theta_{0},\theta_{b}\right)  $, that
is, there exists the unique solution to the system of linear equations:
$\left(  \theta_{0},\theta_{b}\right)  \Re=0$ and $\theta_{0}e+\theta_{b}e=1$.
This completes the proof. \textbf{{\rule{0.08in}{0.08in}}}

The following theorem provides a necessary and sufficient conditions under
which the Markov process $\mathbf{Q}$ is positive recurrence.

\begin{The}
The Markv process $\mathbf{Q}$ of GI/M/1 type is positive recurrent if and
only if
\begin{equation}
\left(  \theta_{0}+\theta_{b}\right)  \left(  D\otimes I\right)  e<b\theta
_{0}\left[  I\otimes\left(  S^{0}\alpha\right)  \right]  e. \label{Equ-2}%
\end{equation}
\end{The}

\textbf{Proof: }By using the mean drift method given in Chapter 3 of Li [18]
(e.g., Theorem 3.19 and the continuous-time case in Page 172), it is easy to
see that the Markv process $\mathbf{Q}$ of GI/M/1 type is positive recurrent
if and only if
\begin{equation}
\theta A_{0}e<b\theta A_{b+1}e. \label{eq-3}%
\end{equation}
Note that
\begin{align}
\theta A_{0}e  &  =\theta_{0}\left(  D\otimes I\right)  e+\theta_{b}\left(
D\otimes I\right)  e\nonumber\\
&  =\left(  \theta_{0}+\theta_{b}\right)  \left(  D\otimes I\right)  e
\label{eq-4}%
\end{align}
and
\begin{equation}
b\theta A_{b+1}e=b\theta_{0}\left[  I\otimes\left(  S^{0}\alpha\right)
\right]  e, \label{eq-5}%
\end{equation}
thus we obtain
\[
\left(  \theta_{0}+\theta_{b}\right)  \left(  D\otimes I\right)  e<b\theta
_{0}\left[  I\otimes\left(  S^{0}\alpha\right)  \right]  e.
\]
This completes the proof. \textbf{{\rule{0.08in}{0.08in}}}

It is necessary to consider a special case in which the transaction inputs are
Poisson with arrival rate $\lambda$, and the blockchain-building and
block-generation times are exponential with service rates $\mu_{1}$ and
$\mu_{2}$, respectively. Note that this special case was studied in Li et al.
\cite{Li:2018}, here we only restate the stable condition as the following corollary.

\begin{Cor}
The Markov process $\mathbf{Q}$ of GI/M/1 type is positive recurrent if and
only if
\begin{equation}
\frac{b\mu_{1}\mu_{2}}{\mu_{1}+\mu_{2}}>\lambda. \label{Equ-3}%
\end{equation}
\end{Cor}

By observing (\ref{Equ-3}), it is easy to see that $1/\left(  b\mu_{1}\right)
+$ $1/\left(  b\mu_{2}\right)  <1/\lambda$, that is, the complicated service
speed of transactions is faster than the transaction arrival speed, under
which the Markov process $\mathbf{Q}$ of GI/M/1 type is positive recurrent.
However, it is not easy to understood from (\ref{Equ-2}).

If the Markv process $\mathbf{Q}$ of GI/M/1 type is positive recurrent, we
write its stationary probability vector as
\[
\mathbf{\pi}=\left(  \mathbf{\pi}_{0},\mathbf{\pi}_{1},\mathbf{\pi}_{2}%
,\ldots\right)  ,
\]
where for $k=0$%
\[
\mathbf{\pi}_{0}=\left(  \pi_{0,0},\pi_{0,1},\ldots,\pi_{0,b}\right)  ,
\]%
\[
\pi_{0,0}=\left(  \pi_{0,0}^{\left(  i\right)  }:1\leq i\leq m_{0}\right)  ,
\]
and for $1\leq l\leq b$%
\[
\pi_{0,l}=\left(  \pi_{0,l}^{\left(  i,j\right)  }:1\leq i\leq m_{0},1\leq
j\leq m_{1}\right)  ;
\]
for $k\geq1$%
\[
\mathbf{\pi}_{k}=\left(  \pi_{k,0},\pi_{k,1},\ldots,\pi_{k,b}\right)  ,
\]%
\[
\pi_{k,0}=\left(  \pi_{k,0}^{\left(  i,r\right)  }:1\leq i\leq m_{0},1\leq
r\leq m_{2}\right)  ,
\]
and for $1\leq l\leq b$
\[
\pi_{k,l}=\left(  \pi_{k,l}^{\left(  i,j\right)  }:1\leq i\leq m_{0},1\leq
j\leq m_{1}\right)  .
\]
Note that in the above expressions, the vector $\mathbf{a}=\left(  a^{\left(
i,j\right)  }:1\leq i\leq I,1\leq j\leq J\right)  $ is based on the
lexicographical order of the elements, that is,%
\[
\mathbf{a}=\left(  a^{\left(  1,1\right)  },a^{\left(  1,2\right)  }%
,\ldots,a^{\left(  1,J\right)  };a^{\left(  2,1\right)  },a^{\left(
2,2\right)  },\ldots,a^{\left(  2,J\right)  };\ldots;a^{\left(  I,1\right)
},a^{\left(  I,2\right)  },\ldots,a^{\left(  I,J\right)  }\right)  .
\]

If $\left(  \theta_{0}+\theta_{b}\right)  \left(  D\otimes I\right)
e<b\theta_{0}\left[  I\otimes\left(  S^{0}\alpha\right)  \right]  e$, then the
Markv process $\mathbf{Q}$ of GI/M/1 type is irreducible and positive
recurrent. Thus the Markov process $\mathbf{Q}$ exists a unique stationary
probability vector, which is also matrix-geometric. Thus, to express the
matrix-geometric stationary probability vector, we need to first obtain the
rate matrix $R$, which is the minimal nonnegative solution to the following
nonlinear matrix equation
\begin{equation}
R^{b+1}A_{b+1}+RA_{1}+A_{0}=0. \label{eq-6}%
\end{equation}

In general, it is very complicated to solve this nonlinear matrix equation
(\ref{eq-6}) due to the term $R^{b+1}A_{b+1}$ of size $b+1$. In fact, for the
blockchain queueing system, here we can not provide an explicit expression for
the rate matrix $R$ yet. In this case, we can use some iterative algorithms,
given in Neuts \cite{Neu:1981}, to give its numerical solution. For example,
an effective iterative algorithm given in Neuts \cite{Neu:1981} is described
as
\begin{align*}
R_{0}  &  =0\mathbf{,}\\
R_{N+1}  &  =\left(  R_{N}^{b+1}A_{b+1}+A_{0}\right)  \left(  -A_{1}\right)
^{-1}.
\end{align*}
Note that this algorithm is fast convergent, that is, after a finite number of
iterative steps, we can numerically obtain a solution of higher precision
which is used to approximate the rate matrix $R$.

The following theorem directly comes from Theorem 1.2.1 of Chapter 1 in Neuts
\cite{Neu:1981}. Here, we restate it without a proof.

\begin{The}
If the Markv process $\mathbf{Q}$ of GI/M/1 type is positive recurrent, then
the stationary probability vector $\mathbf{\pi}=\left(  \mathbf{\pi}%
_{0},\mathbf{\pi}_{1},\mathbf{\pi}_{2},\ldots\right)  $ is given by
\begin{equation}
\mathbf{\pi}_{k}=\mathbf{\pi}_{1}R^{k-1},\ \ \ k\geq2. \label{eq-8}%
\end{equation}
where the vector $\left(  \mathbf{\pi}_{0},\mathbf{\pi}_{1}\right)  $ is the
stationary probability vector of the censoring Markov process $\mathbf{Q}%
^{\left(  1,2\right)  }$ of levels $0$ and $1$ which is irreducible and
positive recurrent. Thus it is the unique solution to the following system of
linear equations:%
\begin{equation}
\left\{
\begin{array}
[c]{c}%
\left(  \mathbf{\pi}_{0},\mathbf{\pi}_{1}\right)  \mathbf{Q}^{\left(
1,2\right)  }=\left(  \mathbf{\pi}_{0},\mathbf{\pi}_{1}\right)  ,\\
\mathbf{\pi}_{0}e+\mathbf{\pi}_{1}\left(  I-R\right)  ^{-1}e=1,
\end{array}
\right.  \label{Equa-4}%
\end{equation}
where%
\[
\mathbf{Q}^{\left(  1,2\right)  }=\left(
\begin{array}
[c]{cc}%
B_{1} & B_{0}\\
\sum\limits_{k=2}^{b+1}R^{k-2}B_{k} & A_{1}+R^{b}A_{b+1}%
\end{array}
\right)  .
\]
\end{The}

\textbf{Proof:} Here, we only derive the boundary condition (\ref{Equa-4}). It
follows from $\mathbf{\pi Q}=0$ that%
\[
\left\{
\begin{array}
[c]{l}%
\pi_{0}B_{1}+\pi_{1}B_{2}+\cdots+\pi_{b}B_{b+1}=0,\\
\pi_{0}B_{0}+\pi_{1}A_{1}+\pi_{b+1}A_{b+1}=0.
\end{array}
\right.
\]
By using the matrix-geometric solution $\mathbf{\pi}_{k}=\mathbf{\pi}%
_{1}R^{k-1}$ for\ $k\geq2$, we have%
\[
\left\{
\begin{array}
[c]{l}%
\pi_{0}B_{1}+\pi_{1}\left(  B_{2}+RB_{3}+\cdots+R^{b-1}B_{b+1}\right)  =0,\\
\pi_{0}B_{0}+\pi_{1}\left(  A_{1}+R^{b}A_{b+1}\right)  =0.
\end{array}
\right.
\]
This gives the desired result, and completes the proof.
\textbf{{\rule{0.08in}{0.08in}}}

\section{The Stationary Transaction Numbers}

In this section, we discuss two key performance measures: The average
stationary numbers of transactions both in the queueing waiting room and in
the block, and give their simple expressions by means of the vectors
$\mathbf{\pi}_{0}$ and $\mathbf{\pi}_{1}$, and the rate matrix $R$. Finally,
we use numerical examples to verify computability of our theoretical results,
and show how the performance measures depend on the main parameters of this system.

If $\left(  \theta_{0}+\theta_{b}\right)  \left(  D\otimes I\right)
e<b\theta_{0}\left[  I\otimes\left(  S^{0}\alpha\right)  \right]  e$, then the
blockchain system is stable. In this case, we write that w.p.1,
\[
N_{1}=lim_{t\rightarrow+\infty}N_{1}\left(  t\right)  ,\ \ \ N_{2}%
=lim_{t\rightarrow+\infty}N_{2}\left(  t\right)  ,
\]
where $N_{1}\left(  t\right)  $ and $N_{2}\left(  t\right)  $ are the numbers
of transactions in the queueing waiting room and of transactions in the block
at time $t\geq0$, respectively.

\textbf{(a) The average stationary number of transactions in the queueing
waiting room}

It follows from (\ref{eq-8}) and (\ref{Equa-4}) that
\begin{align*}
E\left[  N_{1}\right]   &  =\overset{\infty}{\underset{k=1}{\sum}}%
k\overset{m_{0}}{\underset{i=1}{\sum}}\overset{m_{2}}{\underset{r=1}{\sum}}%
\pi_{k,0}^{\left(  i,r\right)  }+\overset{\infty}{\underset{k=1}{\sum}%
}k\underset{l=1}{\overset{b}{\sum}}\overset{m_{0}}{\underset{i=1}{\sum}%
}\overset{m_{1}}{\underset{j=1}{\sum}}\pi_{k,l}^{\left(  i,j\right)  }\\
&  =\overset{\infty}{\underset{k=1}{\sum}}k\underset{l=0}{\overset{b}{\sum}%
}\mathbf{\pi}_{k,l}\text{ }e\\
&  =\overset{\infty}{\underset{k=1}{\sum}}k\text{ }\mathbf{\pi}_{k}\text{
}e=\mathbf{\pi}_{1}R\left(  I-R\right)  ^{-2}e.
\end{align*}
Note that the above three vectors $e$ has different sizes, for example, the
size of the first one is $m_{0}\times m_{2}$ for $l=0$ and $m_{0}\times m_{1}$
for $1\leq l\leq b$; while the sizes of the second and third are $m_{0}%
\times\left(  m_{2}+bm_{1}\right)  $. For simplicity of description, here we
use only a vector $e$ whose size can easily be inferred by the context.

\textbf{(b)} \textbf{The average stationary number of transactions in the block}

Let $\mathbf{h}=\left(  0,e,2e,\ldots,be\right)  ^{T}$. Then
\begin{align*}
E\left[  N_{2}\right]   &  =\underset{l=0}{\overset{b}{\sum}}l\overset{\infty
}{\underset{k=0}{\sum}}\overset{m_{0}}{\underset{i=1}{\sum}}\overset{m_{1}%
}{\underset{j=1}{\sum}}\pi_{k,l}^{\left(  i,j\right)  }\\
&  =\underset{l=0}{\overset{b}{\sum}}l\overset{\infty}{\underset{k=0}{\sum}%
}\mathbf{\pi}_{k,l}\text{ }e\\
&  =\overset{\infty}{\underset{k=0}{\sum}}\mathbf{\pi}_{k}\text{ }\mathbf{h}\\
&  =\left[  \mathbf{\mathbf{\pi}_{0}+\pi}_{1}\left(  I-R\right)  ^{-1}\right]
\text{ }\mathbf{h}.
\end{align*}

In the remainder of this section, we provide some numerical examples to verify
computability of our theoretical results, and to analyze how the two
performance measures $E\left[  N_{1}\right]  $ and $E\left[  N_{2}\right]  $
depend on some crucial parameters of the blockchain queueing system.

In the two numerical examples, we take some common parameters: The
block-building service rate $\mu_{1}\in\left[  0.05,1.5\right]  $, the
block-generation service rate $\mu_{2}=2$, the arrival rate $\lambda=0.3$, the
maximum block size $b=40,320,1000$, respectively.

From Figure 3, it is seen that $E\left[  N_{1}\right]  $ and $E\left[
N_{2}\right]  $ decrease, as $\mu_{1}$ increases. At the same time, $E\left[
N_{1}\right]  $ decreases as $b$ increases, but $E\left[  N_{2}\right]  $
increases as $b$ increases.

\begin{figure}[th]
\centering                 \includegraphics[width=7.3cm]{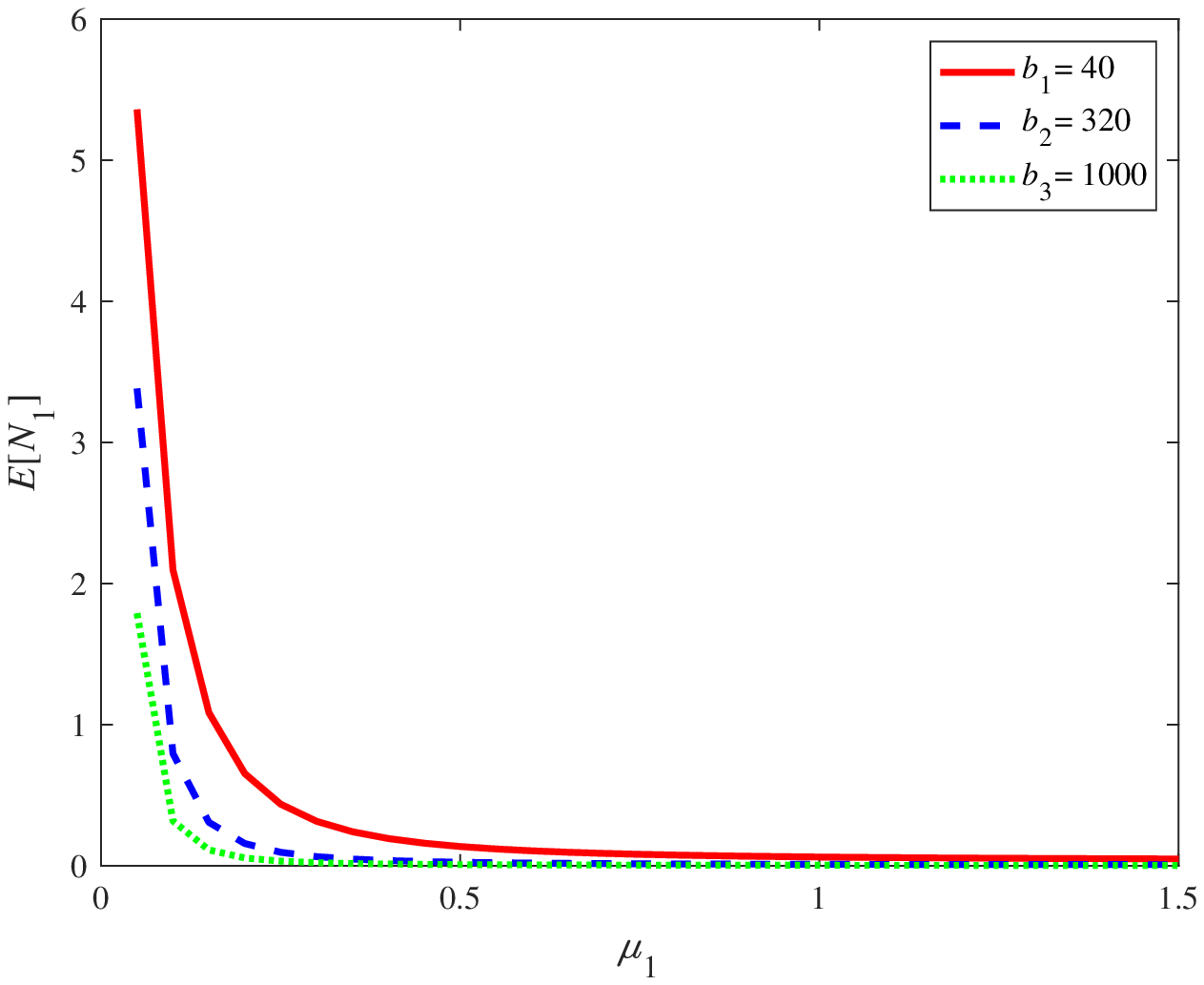}
\centering   \includegraphics[width=7.3cm]{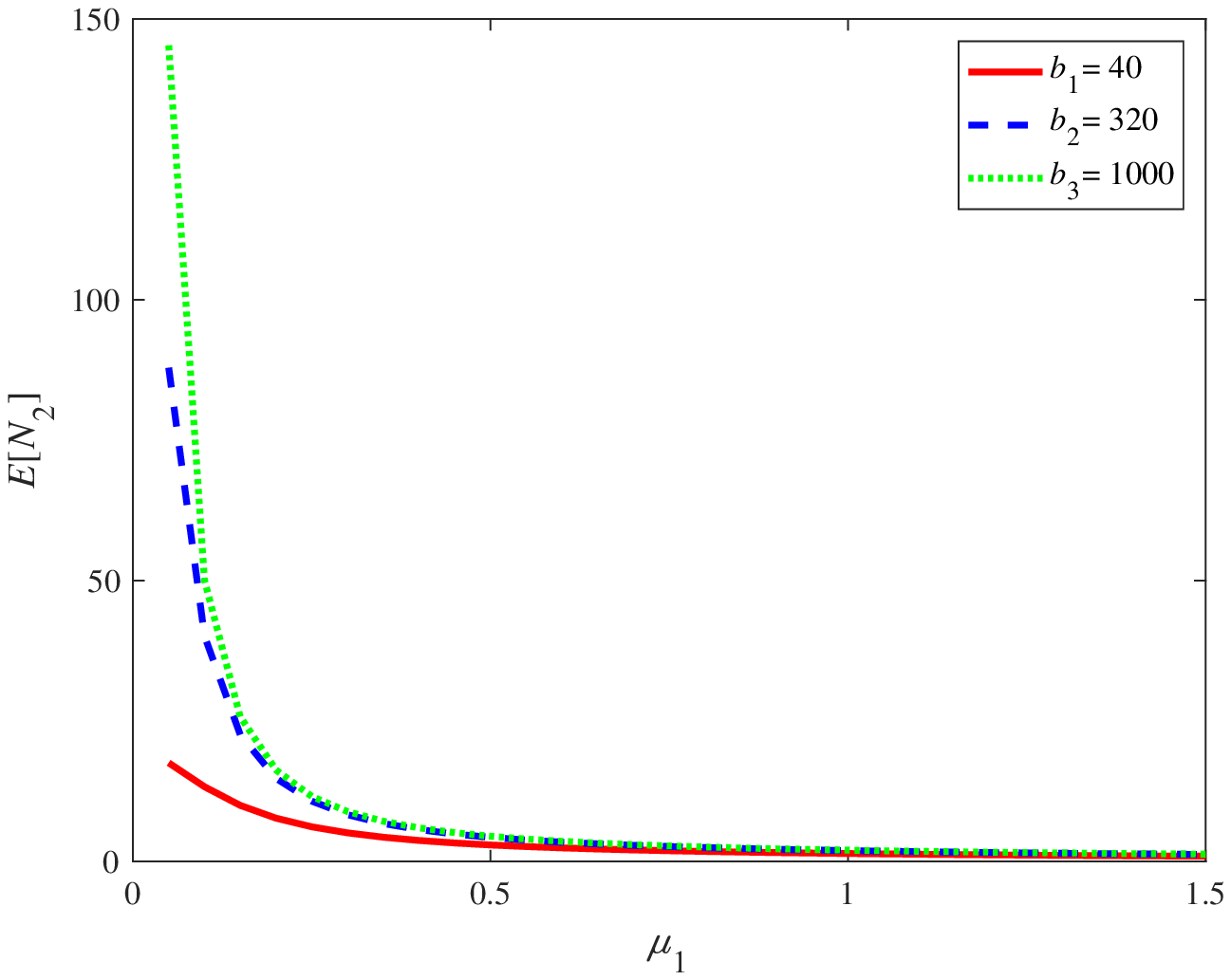}  \caption{$E\left[
N_{1}\right]  $, $E\left[  N_{2}\right]  $ vs. $\mu_{1}$ under three different
values of $b$}%
\label{figure:Fig-3}%
\end{figure}

\section{The Transaction-Confirmation Time}

In this section, we provide a matrix-analytic method based on the
$RG$-factorizations for computing the average transaction-confirmation time of
any transaction, which is always an interesting but difficult topic because of
the batch service for a block of transactions, and the Service-In-Random-Order
for choosing some transactions into a block.

In the blockchain system, the transaction-confirmation time is the time
interval from the time epoch that a transaction arrives at the queueing
waiting room to the time point that the block including the transaction is
first confirmed and then it is built in the blockchain. Obviously, the
transaction-confirmation time is the sojourn time of the transaction in the
blockchain system, and it is the sum of the block-generation and
blockchain-building times of a transaction taken in the block. Let $\Im$
denote the transaction-confirmation time of any transaction when the
blockchain system is stable.

To study the transaction-confirmation time $\Im$, we need to introduce the
stationary life time $\Gamma_{s}$ of the PH blockchain-building time $\Gamma$
with an irreducible representation $\left(  \alpha,T\right)  $. Let $\varpi$
be the stationary probability vector of the Markov process $T+T^{0}\alpha$.
Then the stationary life time $\Gamma_{s}$ is also a PH distribution with an
irreducible representation $\left(  \varpi,T\right)  $, e.g., see Property 1.5
in Chapter 1 of Li \cite{Li:2010}. Clearly, $E\left[  \Gamma_{s}\right]
=-\varpi T^{-1}e.$

Now, we introduce a Markov process $\left\{  Y\left(  t\right)  :t\geq
0\right\}  $ with an absorbing state, whose state transition relation given in
Figure 4 according to Figures 1 and 2. At the same time, we define the first
passage time as%
\[
\xi=\inf\left\{  t:Y\left(  t\right)  =\text{the absorbing state, }%
t\geq0\right\}  .
\]
For $k\geq0,1\leq i\leq m_{0}$ and $1\leq r\leq m_{2}$, if $Y\left(  0\right)
=$ $\left(  k,0;i,r\right)  $, then we write the first passage time as
$\xi_{|\left(  k,0;i,r\right)  }$.

\begin{figure}[th]
\centering                        \includegraphics[width=13cm]{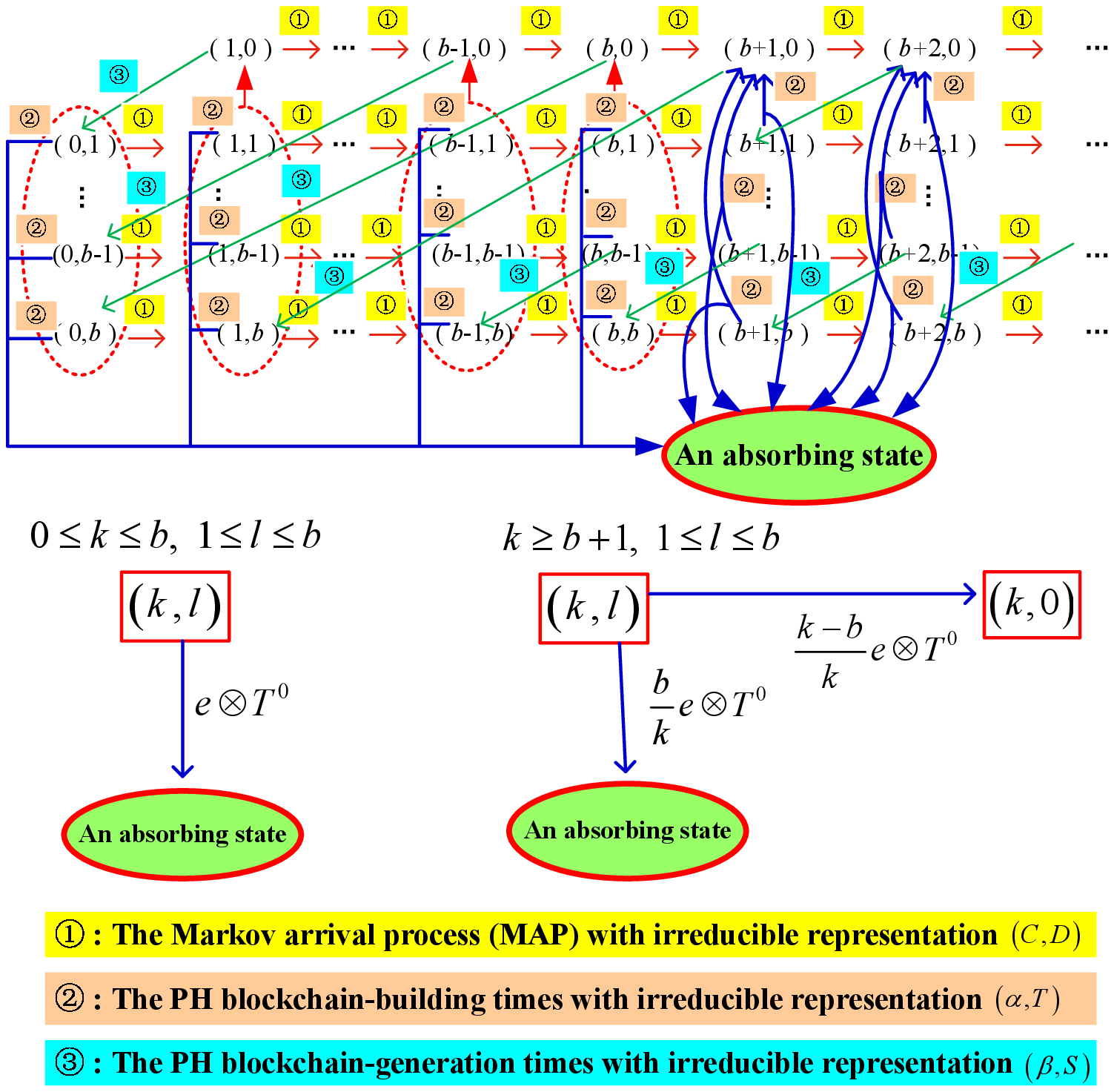}
\newline \caption{State transition relation of the Markov process with an
absorbing state}%
\label{figure:Fig-4}%
\end{figure}

\begin{Rem}
It is necessary to explain the absorbing rates in the below part of Figure 4.

(1) If $Y\left(  0\right)  =\left(  k,l\right)  $ for $1\leq k\leq b$ and
$0\leq l\leq b$, then the $k$ transactions can be chosen into a block once the
previous block is pegged to the blockchain, a tagged transaction of the $k$
transactions is chosen into the block with probability $1$.

(2) If $Y\left(  0\right)  =\left(  k,l\right)  $ for $k\geq b+1$ and $0\leq
l\leq b$, then any $b$ transactions of the $k$ transactions can randomly be
chosen into a block once the previous block is pegged to the blockchain, thus
a tagged transaction of the $k$ transactions is chosen into the block of the
maximal size $b$ with probability $b/k$.
\end{Rem}

When a transaction arrives at the queueing waiting room, it can observes the
states of the blockchain system having two different cases:

\textbf{Case one:} State $\left(  k,0;i,r\right)  $ for $k\geq1;1\leq i\leq
m_{0}$ and $1\leq r\leq m_{2}$. In this case, with the initial probability
$\pi_{k,0}^{\left(  i,r\right)  }$, the transaction-confirmation time $\Im$ is
the first passage time $\xi_{|\left(  k,0;i,r\right)  }$ of the Markov process
with an absorbing state, whose state transition relation given in Figure 4.

\textbf{Case two:} State $\left(  k,l;i,r\right)  $ for $k\geq1,1\leq l\leq
b;1\leq i\leq m_{0}$ and $1\leq j\leq m_{1}$. In this case, with the initial
probability $\pi_{k,l}^{\left(  i,j\right)  }$, the transaction-confirmation
time $\Im$ is decomposed into the sum of the random variable $\Gamma_{s}$ and
the first passage time $\xi_{|\left(  k,0;i,r\right)  }$ given in \textbf{Case
one}. It is easy to see Figure 4 that there exists a stochastic decomposition:
$\Im=\Gamma_{s}+\xi_{|\left(  k,0;i,r\right)  }$.

For the above analysis, it is easy to see that computation of the first
passage time $\xi_{|\left(  k,0;i,r\right)  }$ is a key in analyzing the
transaction-confirmation time.

Based on the state transition relation given in Figure 4, now we write the
infinitesimal generator of the Markov process $\left\{  Y\left(  t\right)
:t\geq0\right\}  $ as
\begin{equation}
\mathbf{H}=\left(
\begin{array}
[c]{ccccccccc}%
\widetilde{B}_{1} & \widetilde{B}_{0} &  &  &  &  &  &  & \\
\widetilde{B}_{2} & \widetilde{A}_{1} & A_{0} &  &  &  &  &  & \\
\widetilde{B}_{3} &  & \widetilde{A}_{1} & A_{0} &  &  &  &  & \\
\vdots &  &  & \ddots & \ddots &  &  &  & \\
\widetilde{B}_{b+1} &  &  &  & \widetilde{A}_{1} & A_{0} &  &  & \\
& A_{b+1} &  &  &  & \widetilde{A}_{1}^{\left(  b+1\right)  } & A_{0} &  & \\
&  & A_{b+1} &  &  &  & \widetilde{A}_{1}^{\left(  b+2\right)  } & A_{0} & \\
&  &  & \ddots &  &  &  & \ddots & \ddots
\end{array}
\right)  , \label{Equ-1}%
\end{equation}
where
\[
A_{0}=\left(
\begin{array}
[c]{cccc}%
D\otimes I &  &  & \\
& D\otimes I &  & \\
&  & \ddots & \\
&  &  & D\otimes I
\end{array}
\right)  ,\text{ }A_{b+1}=\left(
\begin{array}
[c]{cccc}%
0 & \cdots & 0 & I\otimes\left(  S^{0}\alpha\right) \\
&  &  & \\
&  &  & \\
&  &  &
\end{array}
\right)  ,
\]%
\[
\widetilde{A}_{1}=\left(
\begin{array}
[c]{cccc}%
C\oplus S &  &  & \\
& C\oplus T &  & \\
&  & \ddots & \\
&  &  & C\oplus T
\end{array}
\right)  ,
\]
for $k\geq b+1$%
\[
\widetilde{A}_{1}^{\left(  k\right)  }=\left(
\begin{array}
[c]{cccc}%
C\oplus S &  &  & \\
I\otimes\left(  \frac{k-b}{k}T^{0}\beta\right)  & C\oplus T &  & \\
\vdots &  & \ddots & \\
I\otimes\left(  \frac{k-b}{k}T^{0}\beta\right)  &  &  & C\oplus T
\end{array}
\right)  ;
\]%
\[
\widetilde{B}_{0}=\left(
\begin{array}
[c]{ccccc}%
0 & D\otimes I &  &  & \\
&  & D\otimes I &  & \\
&  &  & \ddots & \\
&  &  &  & D\otimes I
\end{array}
\right)  ,\text{ }\widetilde{B}_{1}=\left(
\begin{array}
[c]{cccc}%
C\otimes I &  &  & \\
& C\oplus T &  & \\
&  & \ddots & \\
&  &  & C\oplus T
\end{array}
\right)  ,
\]%
\[
\widetilde{B}_{2}=\left(
\begin{array}
[c]{ccccc}%
I\otimes\left(  S^{0}\alpha\right)  & 0 & 0 & \cdots & 0\\
&  &  &  & \\
&  &  &  & \\
&  &  &  &
\end{array}
\right)  ,\text{ \ }\ldots,\text{ \ }\widetilde{B}_{b+1}=\left(
\begin{array}
[c]{cccc}%
0 & \cdots & 0 & I\otimes\left(  S^{0}\alpha\right) \\
&  &  & \\
&  &  & \\
&  &  &
\end{array}
\right)  .
\]

If the blockchain system is stable, then the probability that a transaction
observes State $\left(  0,0;i\right)  $ only after arrived at the instant is
$\pi_{1,0}^{\left(  i,r\right)  }$; for $1\leq l\leq b$, the probability that
a transaction observes State $\left(  0,l;i,j\right)  $ only after arrived at
the instant is $\pi_{1,l}^{\left(  i,j\right)  }$; for $k\geq2$, the
probability that a transaction observes State $\left(  k-1,0;i,r\right)  $
only after arrived at the instant is $\pi_{k,0}^{\left(  i,r\right)  }$; for
$k\geq2,1\leq l\leq b$, the probability that a transaction observes State
$\left(  k-1,l;i,j\right)  $ only after arrived at the instant is $\pi
_{k,l}^{\left(  i,j\right)  }$. Obviously, for $0\leq l\leq b$, States
$\left(  0,0;i\right)  $ and $\left(  0,l;i,j\right)  $ will not be
encountered by the transaction only after arrived at the instant, thus the
stationary probabilities $\pi_{0,0}^{\left(  i\right)  }$ and $\pi
_{0,l}^{\left(  i,j\right)  }$ should be omitted by the observation of any
arriving transaction. Based on this, we introduce a new initial probability
vector for the observation of any transaction only after arrived at the
instant as follows:
\[
\mathbf{\gamma}=\left(  \mathbf{\gamma}_{1},\mathbf{\gamma}_{2},\mathbf{\gamma
}_{3},\ldots\right)  ,
\]
where for $k\geq1$%
\[
\mathbf{\gamma}_{k}=\left(  \gamma_{k,0},\gamma_{k,1},\ldots,\gamma
_{k,b}\right)  ,
\]%
\[
\gamma_{k,0}=\left(  \frac{1}{1-\mathbf{\pi}_{0}e}\pi_{k,0}^{\left(
i,r\right)  }:1\leq i\leq m_{0},1\leq r\leq m_{2}\right)
\]
and for $1\leq l\leq b$
\[
\gamma_{k,l}=\left(  \frac{1}{1-\mathbf{\pi}_{0}e}\pi_{k,l}^{\left(
i,j\right)  }:1\leq i\leq m_{0},1\leq j\leq m_{1}\right)  .
\]
To emphasize on the event that the transaction observes State $\left(
k-1,0;i,r\right)  $ only after arrived at the instant, we introduce a new
initial probability vector
\[
\mathbf{\varphi}=\left(  \mathbf{\varphi}_{1},\mathbf{\varphi}_{2}%
,\mathbf{\varphi}_{3},\ldots\right)  ,
\]
where for $k\geq1$%
\[
\mathbf{\varphi}_{k}=\left(  \gamma_{k,0},0,0,\ldots,0\right)  .
\]
In addition, we take%
\[
\mathbf{\psi}=\mathbf{\gamma-\varphi.}%
\]

\begin{The}
\label{The:PH-1}If the blockchain system is stable, then the first passage
time $\xi_{|\left(  k,0;i,r\right)  }$ is a PH distribution of infinite size
with an irreducible representation $\left(  \eta\left(  k,0;i,r\right)
,\mathbf{H}\right)  $, where $\mathbf{H}$ is given in (\ref{Equ-1}), and%
\[
\eta\left(  k,0;i,r\right)  =\left(  0,0,\ldots,0,\frac{1}{1-\mathbf{\pi}%
_{0}e}\pi_{k,0}^{\left(  i,r\right)  },0,0,\ldots,0\right)  .
\]
Also, we have%
\begin{align*}
\mathbf{H}^{0}  &  =-\mathbf{H}e\\
&  =\left(  e\otimes T^{0},e\otimes T^{0},\ldots,e\otimes T^{0};\frac{b}%
{b+1}e\otimes T^{0},\frac{b}{b+2}e\otimes T^{0},\ldots\right)  .
\end{align*}
\end{The}

\textbf{Proof:} If the blockchain system is stable, then $\xi_{|\left(
k,0;i,r\right)  }$ is the first passage time of the Markov process
$\mathbf{H}$ (or $\left\{  Y\left(  t\right)  :t\geq0\right\}  $) with an
absorbing state and under the initial state $Y\left(  0\right)  =$ $\left(
k,0;i,r\right)  $. Note that the original Markov process $\mathbf{Q}$ given in
(\ref{eq-2}) is irreducible and positive recurrent, thus $\xi_{|\left(
k,0;i,r\right)  }$ is a PH distribution of infinite size with an irreducible
representation $\left(  \eta\left(  k,0;i,r\right)  ,\mathbf{H}\right)  $. At
the same time, a simple computation gives%
\[
\mathbf{H}^{0}=\left(  e\otimes T^{0},e\otimes T^{0},\ldots,e\otimes
T^{0};\frac{b}{b+1}e\otimes T^{0},\frac{b}{b+2}e\otimes T^{0},\ldots\right)
.
\]
This completes the proof. \textbf{{\rule{0.08in}{0.08in}}}

Based on Theorem \ref{The:PH-1}, now we extend the first passage time
$\xi_{|\left(  k,0;i,r\right)  }$ to $\xi_{|\left(  \mathbf{0},\mathbf{\varphi
}\right)  }$, which is the first passage time of the Markov process
$\mathbf{H}$ with an initial probability vector $\left(  \mathbf{0}%
,\mathbf{\varphi}\right)  $. The following corollary shows that $\xi_{|\left(
\mathbf{0},\mathbf{\varphi}\right)  }$ is PH distribution of infinite size,
while its proof is easy and is omitted here.

\begin{Cor}
\label{Cor:PH-2}If the blockchain system is stable, then the first passage
time $\xi_{|\left(  \mathbf{0},\mathbf{\varphi}\right)  }$ is a PH
distribution of infinite size with an irreducible representation $\left(
\left(  \mathbf{0},\mathbf{\varphi}\right)  ,\mathbf{H}\right)  $, and%
\[
E\left[  \xi_{|\left(  \mathbf{0},\mathbf{\varphi}\right)  }\right]  =-\left(
\mathbf{0},\mathbf{\varphi}\right)  \mathbf{H}^{-1}e,
\]%
\[
Var\left[  \xi_{|\left(  \mathbf{0},\mathbf{\varphi}\right)  }\right]
=\left(  \mathbf{0},\mathbf{\varphi}\right)  \mathbf{H}^{-2}e-\left[  \left(
\mathbf{0},\mathbf{\varphi}\right)  \mathbf{H}^{-1}e\right]  ^{2}.
\]
\end{Cor}

The following theorem provides a simple expression for the average
transaction-confirmation time $E[\Im]$ by means of Corollary \ref{Cor:PH-2}.

\begin{The}
\label{The:PH-2}If the blockchain queueing system is stable, then the average
transaction-confirmation time $E[\Im]$ is given by
\[
E\left[  \Im\right]  =E\left[  \xi_{|\left(  \mathbf{0},\mathbf{\varphi
}\right)  }\right]  +\left(  1-\mathbf{\varphi}e\right)  E\left[  \Gamma
_{s}\right]  ,
\]
where $\Gamma_{s}$ is the stationary life time of the PH blockchain-building
time with an irreducible representation $\left(  \alpha,T\right)  $. Further,
we have%
\[
E\left[  \Im\right]  =-\left(  \mathbf{0},\mathbf{\varphi}\right)
\mathbf{H}^{-1}e-\left(  1-\mathbf{\varphi}e\right)  \varpi T^{-1}e,
\]
where $\varpi$ is the stationary probability vector of the Markov process
$T+T^{0}\alpha$.
\end{The}

\textbf{Proof:} We first introduce two basic events%
\begin{align*}
\Theta=  &  \left\{  \text{The transaction observes States }\left(
0,0;i\right)  \text{ and }\left(  k,0;i,r\right)  \right.  \text{\ }\\
&  \text{ }\left.  \text{for }1\leq i\leq m_{0},k\geq1,1\leq r\leq
m_{2}\right.  \text{ }\\
&  \left.  \text{only after arrived at the instant}\right\}
\end{align*}
and%
\begin{align*}
\Theta^{c}=  &  \left\{  \text{The transaction observes States }\left(
k,l;i,j\right)  \right.  \text{\ }\\
&  \text{ }\left.  \text{for }k\geq1,1\leq l\leq b,1\leq i\leq m_{0},1\leq
j\leq m_{1}\right.  \text{ }\\
&  \left.  \text{only after arrived at the instant}\right\}  .
\end{align*}
It is easy to see that $\Theta\cup\Theta^{c}=\Omega$. Thus the two events is
complementary according to the fact that the transaction can observe all the
states of the Markov process $\mathbf{Q}$ only after arrived at the instant.
If the blockchain system is stable, then it is easy to compute the
probabilities of the two events as follows:%
\[
P\left\{  \Theta\right\}  =\left(  \mathbf{0},\mathbf{\varphi}\right)
e=\mathbf{\varphi}e
\]
and%
\[
P\left\{  \Theta^{c}\right\}  =1-P\left\{  \Theta\right\}  =1-\mathbf{\varphi
}e.
\]
By using the law of total probability, we obtain%
\begin{align*}
E\left[  \Im\right]   &  =P\left\{  \Theta\right\}  E\left[  \Im\text{
}|\text{ }\Theta\right]  +P\left\{  \Theta^{c}\right\}  E\left[  \Im\text{
}|\text{ }\Theta^{c}\right] \\
&  =\mathbf{\varphi}e\text{ }E\left[  \xi_{|\left(  \mathbf{0},\mathbf{\varphi
}\right)  }\right]  +\left(  1-\mathbf{\varphi}e\right)  E\left[  \Gamma
_{s}+\xi_{|\left(  \mathbf{0},\mathbf{\varphi}\right)  }\right] \\
&  =E\left[  \xi_{|\left(  \mathbf{0},\mathbf{\varphi}\right)  }\right]
+\left(  1-\mathbf{\varphi}e\right)  E\left[  \Gamma_{s}\right] \\
&  =-\left(  \mathbf{0},\mathbf{\varphi}\right)  \mathbf{H}^{-1}e-\left(
1-\mathbf{\varphi}e\right)  \varpi T^{-1}e.
\end{align*}
The proof is completed. \textbf{{\rule{0.08in}{0.08in}}}

As shown in Theorem \ref{The:PH-2}, it is a key in the study of PH
distributions of infinite sizes whether or not we can compute the inverse
matrix $\mathbf{H}^{-1}$ of infinite size. To this end, we need to use the
$RG$-factorizations, given in Li \cite{Li:2010}, to provide such a computable
path. In what follows we provide only a simple interpretation on the
computation, while some detailed discussions will be left in our another paper
in the future.

In fact, it is often very difficult and challenging to compute the inverse of
a matrix of infinite size only except for the triangular matrices.
Fortunately, by using the $RG$-factorizations, the infinitesimal generator
$\mathbf{H}$ can be decomposed into a product of three matrices: Two
block-triangular matrices and a block-diagonal matrix. Therefore, the
$RG$-factorizations play a key role to generalizing the PH distributions from
finite dimensions to infinite dimensions.

By using Subsection 2.2.3 in Chapter 2 of Li \cite{Li:2010} (see Pages 88 to
89), now we provide the UL-type $RG$-factorization of the infinitesimal
generator $\mathbf{H}$. It will be seen that the $RG$-factorization of
$\mathbf{H}$\ has a beautiful block structure, which is well related to the
special block characteristics of $\mathbf{H}$ corresponding to the blockchain
system. To this end, we need to define and compute the $R$-, $U$- and
$G$-measures as follows.

\textbf{The }$R$\textbf{-measure.} Let $R_{k}$ for $k\geq0$ be the minimal
nonnegative solution to the system of nonlinear matrix equations:%
\begin{align*}
R_{0}  &  =\widetilde{B}_{0}+R_{0}\widetilde{A}_{1}+R_{0}R_{1}\cdots
R_{b-1}R_{b}A_{b+1},\\
R_{1}  &  =A_{0}+R_{1}\widetilde{A}_{1}+R_{1}R_{2}\cdots R_{b}R_{b+1}%
A_{b+1},\\
R_{2}  &  =A_{0}+R_{2}\widetilde{A}_{1}+R_{2}R_{3}\cdots R_{b+1}R_{b+2}%
A_{b+1},\\
&  \vdots\\
R_{b-1}  &  =A_{0}+R_{b-1}\widetilde{A}_{1}+R_{b-1}R_{b}\cdots R_{2b-2}%
R_{2b-1}A_{b+1},
\end{align*}
and%
\begin{align*}
R_{b}  &  =A_{0}+R_{b}\widetilde{A}_{1}^{\left(  b+1\right)  }+R_{b}%
R_{b+1}\cdots R_{2b-1}R_{2b}A_{b+1},\\
R_{b+1}  &  =A_{0}+R_{b+1}\widetilde{A}_{1}^{\left(  b+2\right)  }%
+R_{b+1}R_{b+2}\cdots R_{2b}R_{2b+1}A_{b+1},\\
R_{b+2}  &  =A_{0}+R_{b+2}\widetilde{A}_{1}^{\left(  b+3\right)  }%
+R_{b+2}R_{b+3}\cdots R_{2b+1}R_{2b+2}A_{b+1},\\
&  \vdots
\end{align*}

\textbf{The }$U$\textbf{-measure.} Based on the $R$-measure $R_{k}$ for
$k\geq0$, we have%
\begin{align*}
U_{0}  &  =\widetilde{B}_{1}+R_{0}\widetilde{B}_{2}+R_{0}R_{1}\widetilde
{B}_{3}+\cdots+R_{0}R_{1}\cdots R_{b-2}R_{b-1}\widetilde{B}_{b+1},\\
U_{1}  &  =\widetilde{A}_{1}+R_{1}R_{2}\cdots R_{b-1}R_{b}A_{b+1},\\
U_{2}  &  =\widetilde{A}_{1}+R_{2}R_{3}\cdots R_{b}R_{b+1}A_{b+1},\\
&  \vdots\\
U_{b}  &  =\widetilde{A}_{1}+R_{b}R_{b+1}\cdots R_{2b-2}R_{2b-1}A_{b+1},
\end{align*}
and%
\begin{align*}
U_{b+1}  &  =\widetilde{A}_{1}^{\left(  b+1\right)  }+R_{b+1}R_{b+2}\cdots
R_{2b-1}R_{2b}A_{b+1},\\
U_{b+2}  &  =\widetilde{A}_{1}^{\left(  b+2\right)  }+R_{b+2}R_{b+3}\cdots
R_{2b}R_{2b+1}A_{b+1},\\
U_{b+3}  &  =\widetilde{A}_{1}^{\left(  b+3\right)  }+R_{b+3}R_{b+4}\cdots
R_{2b+1}R_{2b+2}A_{b+1},\\
&  \vdots
\end{align*}

\textbf{The }$G$\textbf{-measure.} Based on the $R$-measure $R_{k}$ for
$k\geq0$ and the $U$-measure $U_{k}$ for $k\geq0$, we have%
\begin{align*}
G_{1,0}  &  =\left(  -U_{1}\right)  ^{-1}\left(  \widetilde{B}_{2}%
+R_{1}\widetilde{B}_{3}+R_{1}R_{2}\widetilde{B}_{4}+\cdots+R_{1}R_{2}\cdots
R_{b-2}R_{b-1}\widetilde{B}_{b+1}\right)  ,\\
G_{2,0}  &  =\left(  -U_{2}\right)  ^{-1}\left(  \widetilde{B}_{3}%
+R_{2}\widetilde{B}_{4}+R_{2}R_{3}\widetilde{B}_{5}+\cdots+R_{2}R_{3}\cdots
R_{b-2}R_{b-1}\widetilde{B}_{b+1}\right)  ,\\
&  \vdots\\
G_{b-1,0}  &  =\left(  -U_{b-1}\right)  ^{-1}\left(  \widetilde{B}_{b}%
+R_{b-1}\widetilde{B}_{b+1}\right)  ,\\
G_{b,0}  &  =\left(  -U_{b}\right)  ^{-1}\widetilde{B}_{b+1},
\end{align*}%
\begin{align*}
G_{2,1}  &  =\left(  -U_{2}\right)  ^{-1}R_{2}R_{3}\cdots R_{b-1}R_{b}%
A_{b+1},\\
G_{3,1}  &  =\left(  -U_{3}\right)  ^{-1}R_{3}R_{4}\cdots R_{b-1}R_{b}%
A_{b+1},\\
&  \vdots\\
G_{b,1}  &  =\left(  -U_{b}\right)  ^{-1}R_{b}A_{b+1},\\
G_{b+1,1}  &  =\left(  -U_{b+1}\right)  ^{-1}A_{b+1},
\end{align*}
and for $k\geq3$%
\begin{align*}
G_{k,k-1}  &  =\left(  -U_{k}\right)  ^{-1}R_{k}R_{k+1}\cdots R_{k+b-3}%
R_{k+b-2}A_{b+1},\\
G_{k+1,k-1}  &  =\left(  -U_{k+1}\right)  ^{-1}R_{k+1}R_{k+2}\cdots
R_{k+b-3}R_{k+b-2}A_{b+1},\\
&  \vdots\\
G_{k+b-2,k-1}  &  =\left(  -U_{k+b-2}\right)  ^{-1}R_{k+b-2}A_{b+1},\\
G_{k+b-1,k-1}  &  =\left(  -U_{k+b-1}\right)  ^{-1}A_{b+1}.
\end{align*}

Based on the $R$-, $U$- and $G$-measures, we provide the UL-type
$RG$-factorization of the infinitesimal generator $\mathbf{H}$ as follows:%
\[
\mathbf{H=}\left(  I-\mathbf{R}_{U}\right)  \mathbf{U}\left(  I-\mathbf{G}%
_{L}\right)  ,
\]
where%
\[
\mathbf{R}_{U}=\left(
\begin{array}
[c]{cccccc}%
0 & R_{0} &  &  &  & \\
& 0 & R_{1} &  &  & \\
&  & 0 & R_{2} &  & \\
&  &  & 0 & R_{3} & \\
&  &  &  & \ddots & \ddots
\end{array}
\right)  ,
\]%
\[
\mathbf{U}=\text{diag}\left(  U_{0},U_{1},U_{2},U_{3},\ldots\right)
\]
and%
\[
\mathbf{G}_{L}=\left(
\begin{array}
[c]{ccccccccc}%
0 &  &  &  &  &  &  &  & \\
G_{1,0} & 0 &  &  &  &  &  &  & \\
G_{2,0} & G_{2,1} & 0 &  &  &  &  &  & \\
\vdots & \vdots & \vdots & \ddots &  &  &  &  & \\
G_{b-1,0} & G_{b-1,1} & G_{b-1,b-2} & \cdots & 0 &  &  &  & \\
G_{b,0} & G_{b,1} & G_{b,b-2} & \cdots & G_{b,k} & 0 &  &  & \\
& G_{b+1,1} & G_{b+1,b-2} & \cdots & G_{b+1,k} & G_{b+1,k+1} & 0 &  & \\
&  & G_{b+2,b-2} & \cdots & G_{b+2,k} & G_{b+2,k+1} & G_{b+2,k+2} & 0 & \\
&  &  & \ddots & \vdots & \vdots & \vdots & \vdots & \ddots
\end{array}
\right)  .
\]
Based on the UL-type $RG$-factorization $\mathbf{H=}\left(  I-\mathbf{R}%
_{U}\right)  \mathbf{U}\left(  I-\mathbf{G}_{L}\right)  $, we obtain%
\[
\mathbf{H}^{-1}\mathbf{=}\left(  I-\mathbf{G}_{L}\right)  ^{-1}\mathbf{U}%
^{-1}\left(  I-\mathbf{R}_{U}\right)  ^{-1},
\]
where the inverse matrices $\left(  I-\mathbf{G}_{L}\right)  ^{-1}$,
$\mathbf{U}^{-1}$ and $\left(  I-\mathbf{R}_{U}\right)  ^{-1}$ are given some
expressions in Appendix A.3 of Li \cite{Li:2010}: Inverses of Matrices of
Infinite Size (see Pages 654 to 658). Once the inverse of matrix $\mathbf{H}$
of infinite size is given, the PH distribution of infinite size can be
constructed under a computable and feasible framework. In fact, this is very
important in the study of stochastic models. Also see Li et al. \cite{Li:2005}
and Takine \cite{Tak:2016} for more details.

\begin{Rem}
In general, it is always very difficult and challenging to discuss the
transaction-confirmation time of any transaction in a blockchain system due to
two key points: The block service is a class of batch service, and some
transactions are chosen into a block by means of the Service-In-Random-Order.
For a more general blockchain system, this paper sets up a Markov process with
an absorbing state, and shows that the transaction-confirmation time is the
first passage time of the Markov process with an absorbing state. Therefore,
this paper can discuss the transaction-confirmation time by means of the PH
distribution of infinite size (corresponding to the first passage time), and
provides an effective algorithm for computing the average
transaction-confirmation time by using the $RG$-factorizations of
block-structured Markov processes of infinite levels. We believe that the
$RG$-factorizations of block-structured Markov processes will play a key role
in the queueing study of blockchain systems.
\end{Rem}

\section{Concluding Remarks}

In this paper, we develop a more general framework of block-structured Markov
processes in the queueing study of blockchain systems. To do this, we design a
two stage, Service-In-Random-Order and batch service queueing system with MAP
transaction arrivals and two-stages of PH service times, and discuss some key
performance measures such as the stationary average number of transactions in
the queueing waiting room, the stationary average number of transactions in
the block, and the average transaction-confirmation time of any transaction.
Note that the study of performance measures is a key to improve blockchain
technologies sufficiently. On the other hand, an original aim of this paper is
to generalize the two-stage batch-service queueing model studied in Li et al.
\cite{Li:2018} both ``from exponential to phase-type'' service times and
``from Poisson to MAP'' transaction arrivals. Note that the MAP transaction
arrivals and the two stages of PH service times make our queueing model more
suitable to various practical conditions of blockchain systems with key
factors, for example, the mining processes, the reward incentive, the
consensus mechanism, the block-generation, the blockchain-building and so forth.

By using the matrix-geometric solution, we first obtain a sufficient stable
condition of the blockchain system. Then we provide simple expressions for two
key performance measures: The stationary average number of transactions in the
queueing waiting room, and the stationary average number of transactions in
the block. Finally, to deal with the transaction-confirmation time, we develop
a computational technique of the first passage times by means of both the PH
distributions of infinite sizes and the $RG$-factorizations. In addition, we
use numerical examples to verify computability of our theoretical results.
Along these lines, we will continue our future research on several interesting
directions as follows:

-- Developing effective algorithms for computing the average
transaction-confirmation times in terms of the $RG$-factorizations.

-- Analyzing multiple classes of transactions in the blockchain systems, in
which the transactions are processed in the block-generation and
blockchain-building processes according to a priority service discipline.

-- When the arrivals of transactions are a renewal process, and/or the
block-generation times and/or the blockchain-building times follow general
probability distributions, an interesting future research is to focus on fluid
and diffusion approximations of blockchain systems.

-- Setting up reward function with respect to cost structures, transaction
fees, mining reward, consensus mechanism, security and so forth. It is very
interesting in our future study to develop stochastic optimization, Markov
decision processes and stochastic game models in the study of blockchain systems.

\section*{Acknowledgements}

Q.L. Li was supported by the National Natural Science Foundation of China
under grant No. 71671158, and the Natural Science Foundation of Hebei Province
in China under grant No. G2017203277.

\end{document}